\documentstyle[12pt]{article}
\def\beq{\begin{equation}}
\def\eeq{\end{equation}}
\def\bea{\begin{eqnarray}}
\def\eea{\end{eqnarray}}
\def\bq{\begin{quote}}
\def\eq{\end{quote}}

\def\gappeq{\mathrel{\rlap {\raise.5ex\hbox{$>$}}
{\lower.5ex\hbox{$\sim$}}}}

\def\lappeq{\mathrel{\rlap{\raise.5ex\hbox{$<$}}
{\lower.5ex\hbox{$\sim$}}}}

\def\Toprel#1\over#2{\mathrel{\mathop{#2}\limits^{#1}}}

\begin{document}
\pagestyle{empty}\begin{flushright}
{{\bf CERN-PH-TH-2004-242} \\Revised version, 24 June 2005} \end{flushright}

\vspace*{5mm}
\begin{center}
{\bf BEHAVIOUR OF THE WAVE FUNCTION NEAR THE ORIGIN IN THE RADIAL CASE} \\
\vspace*{1cm} 
{\bf ANDR\'E MARTIN} \\
{\bf Department of Physics, Theory Division \\
CERN\\CH-1211 Geneva 23}\\
\vspace*{2cm}
ABSTRACT  
\end{center}
\vspace*{5mm}
We give one more proof in two and three space dimensions that the irregular solution of the Schr\"odinger equation, for zero angular
momentum, is in fact the solution of an equation containing an extra ``delta function".  We propose another criterium to eliminate the
irregular solution which is to require the validity of the virial theorem of which we give a general proof in the classical and quantum
cases.


\setcounter{page}{1}
\pagestyle{plain}

\section{Introduction}

These pedagogical remarks were stimulated by an exchange of ideas with  Yves
Cantelaube.

We re-examine the following question:  in the case of a central potential why choose the ``regular solution" at the origin of the
Schr\"odinger equation, while the irregular solution may also be square integrable?   It
is indeed easy to show that for any negative energy there exists, under weak conditions on the potential, a solution exponentially
decreasing at infinity which, outside the conventional eigenvalues, is irregular at the origin and square integrable. This is
explicitly demonstrated in Section 2. I would just like first to present, in my own way, the standard argument (Dirac\cite{aaa}, Van
Hove\cite{bb}, Cohen-Tannoudji et al.\cite{cc}, Basdevant et al.\cite{dd}) that the irregular solution is not a solution of the
original Schr\"odinger equation but a solution of an equation containing an extra delta function.  Then, I would like to propose
another criterium:  the violation of the virial theorem by the irregular solution.  I refuse to enter into the questions of
self-adjointness.  You prove that a certain operator is self-adjoint on a certain domain.  Choosing the domain is more or less choosing
the answer.

\section{Proof of the existence of singular, normalizable solutions}

Consider the reduced Schr\"odinger equation in three dimensions for $\ell = 0$:
\bea
\left ( - \frac{d^2}{dr^2} + V(r) - E \right ) u = 0 \nonumber \\
u = r \psi .
\label{one}
\eea
Assume $V \to 0$ for $r \to \infty$.  Then for $E < 0$, there exists $R$ such that
\beq
V(r) - E \geq K^2 > 0 \; {\rm for} \; r \geq R .
\label{2}
\eeq
Consider the solution such that 
$$
u'(R) = 0 .
$$
By comparing Eq. (1) with
\beq
\left ( - \frac{d^2}{dr^2} + K^2 \right ) \; u = 0
\label{3}
\eeq
it is easy to see that
\bea
u(r) > u(R) \; {\rm cos h} K(r-R) \nonumber \\
{\rm for} \; r > R ,
\label{four}
\eea
and that $u(r)$ is increasing.

We can define another solution of Eq. (1), $v$ which has a constant Wronskian with $u$:
\beq
u'v - v'u = 1
\label{five}
\eeq
A particular solution of (5) is 
\beq
v = u \; \int^{\infty}_{r}
\; \frac{dr'}{u^2(r')}
\label{six}
\eeq
From (4) we get
\beq
v < \int^{\infty}_{r} \; 
\frac{dr'}{u(R) {\rm cos h} K(r-R)} = 
\frac{2}{K} \; \frac{{\rm exp} -K(r-R)}{u(R)}
\label{seven}
\eeq
The solution $v$ can be continued for $r < R$ till the origin.

Equation (1) can be written in integral form as
\beq
v = v (R_0) + (r-R_0) v' (R_0) - 
\int_r^{R_0} \; (r'-r) (V(r') - E) v(r') dr'
\label{eight}
\eeq
Hence, for $r < R_0$,
\bea
M(r,R_0) & < & |v(R_0)| + R_0 |v'(R_0)| \nonumber \\
& +  & \int_0^{R_0}  \; r' \left [ |V(r')| + |E| \right ] M(r, R_0) dr' ,  \nonumber 
\eea
where 
\bea
M(r,R_0) & = & {\rm sup} |v(r')| . \nonumber \\
& & r < r' < R_0 \nonumber
\eea 
If
\beq
\int_0^{c} \; r' |V (r')| dr' < \infty ,
\label{nine}
\eeq
we can choose $R_0$ in such a way that
$$
\int_0^{R_0} \; r' [|V(r')| + |E|] dr' < \epsilon < 1
$$
and hence
\beq
M(r,R_0) < 
\frac{|v(R_0)| + R_0 |v'(R_0)|}{1 - \epsilon}
\label{ten}
\eeq
so $|v(r)|$ has a fixed bound, independent of $r$.

Condition (9) is only sufficient.
In particular one can have acceptable oscillating potentials as shown by Chadan and collaborators [5] which violate (9).

Now, since the wave function is given by
$$
\psi = {\rm const.} \; r v ,
$$
it is clear that $\psi$ is normalizable because $\int |\psi|^2 d^3 r$ converges both at the origin and at infinity.  This is valid for
any $E < 0$, in particular for a value $E$ which is \underline{not} a conventional eigenvalue (for which $v$ would tend to zero at the
origin).

\section{The delta function in the radial case with three-dimensions}

We want to prove that the solution of the Schr\"odinger equation
\beq
\psi = \frac{1}{r}~u
\label{eleven}\eeq
cannot behave like $\frac{1}{r}$ at the origin.

 Take
\beq
f =\frac{1}{\sqrt{r^2+a^2}}
\label{two}
\eeq
\beq
\Delta f =\frac{1}{r^2}~\frac{d}{dr}~r^2~\frac{d}{dr}~f = -3~\frac{a^2}{(r^2+a^2)^{5/2}} = -g
\label{twelve}
\eeq
Let us prove that for $a \rightarrow 0$, $g$ tends to $4\pi \delta^3/(\vec r)$.

\begin{itemize}
\item[1)] for $r = 0,~g \rightarrow \infty$ for $a \rightarrow 0$
\item[2)] for any $r > 0, g\rightarrow 0$, for $a \rightarrow 0$
\item[3)] $
\int d^3rg(\vec r) = 4\pi \int^\infty_0 \frac{3a^2r^2dr}{(a^2+r^2)^{5/2}} = 4\pi$. 
\end{itemize}
Hence for 
\bea
a \rightarrow 0, g \rightarrow 4\pi \delta^3(\vec r) \nonumber \\
\Delta_3(-\frac{1}{r}) = +4\pi\delta^3(\vec r) .
\label{thirteen}
\eea

Notice that the \underline{sign}
is \underline{obvious} since $-1/r$ is a subharmonic function.  Hence, as announced the irregular solution of the Schr\"odinger equation is not acceptable.

\section{The case of two-dimensions}

The regular solution for zero azimuthal angular momentum behaves like 1, the irregular solution behaves like $\ln r$.  We use
\beq
 f = \ln \sqrt{r^2+a^2} = \frac{1}{2}~\ln (r^2+a^2)
\label{fourteen}
\eeq
\bea
\Delta f &=& \frac{1}{2} \frac{1}{r}\frac{d}{dr}r\frac{d}{dr}~\ln (r^2+a^2) \nonumber \\
&=& \frac{2a^2}{(r^2+a^2)^2} = g
\label{fifteen}
\eea
Let us prove that $g$ is proportional to a delta function in the limit $a \rightarrow 0$
\begin{itemize}
\item[1)] again $g \rightarrow \infty$ for $r =0~a \rightarrow 0$
\item[2)] $g \rightarrow 0$ for $r > 0~a \rightarrow 0$
\item[3)] 
\beq
\int d^2rg = 2\pi\int rdr \frac{2a^2}{(r^2+a^2)^2} = 2\pi
\label{sixteen}
\eeq
\end{itemize} 
Hence
\beq
g \rightarrow 2\pi \delta^2(\vec r)
\label{seventeen}
\eeq
for $a \rightarrow 0$.  Again the sign is correct because $\ln r$ is \underline{subharmonic} since $\ln r$ tends to MINUS infinity for
$r \rightarrow 0$.

\section{The virial theorem}

The virial theorem should hold classically and ``quantically".  Let me remind the \underline{classical} proof in the general case.
\beq
{\rm If}~~~ \vec F_i =  -\vec\nabla_iV(x_1,x_2,...,x_n)~,
\label{eighteen}
\eeq
Newton's equations are
\beq
m_i~\frac{d^2 \vec x_i}{dt^2} =  -\vec\nabla_iV~.
\label{nineteen}
\eeq
Hence
\beq
\sum m_i \vec x_i \cdot \frac{d^2 \vec x_i}{dt} = - \sum \vec x_i \vec\nabla_iV
\label{twenty}
\eeq
and
\beq
-\int^T_0 \sum m_i \vec x_i \cdot \frac{d^2\vec x_i}{dt} = -\sum m_i \vec x_i \frac{d\vec x_i}{dt} \left|^T_0 + \right.
\int^T_0 \sum m_i \left(\frac{d \vec x_i}{dt}\right)^2 dt 
= \int^T_0 \sum \vec x_i \vec\nabla_i Vdt\label{twentyone}
\eeq
So if you deal with a \underline{confined} system (big problem!)
$$
\lim_{T \rightarrow \infty }~\frac{1}{T}\left(\sum m_i \vec x_i \frac{d\vec x_i}{dt}\right)^T_0 = 0~.\nonumber
$$
Strictly speaking, we know only that the above parenthesis is {\it bounded for almost every} $T$ if the $x_i$'s are bounded.  This
inconvenient could be avoided by smoothing the cut-off at $T$ in the integrals.

With this caveat, the average over \underline{time} of the kinetic energy is:
\beq
\lim_{T \rightarrow \infty}~\frac{1}{T}\int^T_0 \frac{1}{2} \sum m_i \left(\frac{d\vec x_i}{dt}\right)^2 dt = \lim
\frac{1}{T}\int^\infty_0 \frac{(\sum\vec x_i \vec \nabla_i V)}{2} dt~.
\label{twentytwo}
\eeq If some particles escape at infinity with velocities $\vec v_i$ (nobody knows if this will happen in the solar system), we get
\beq
\lim_{T \rightarrow \infty}~\frac{1}{T}~\int^T_0~\frac{1}{2}\sum m_i \left(\frac{d\vec x_i}{dt}\right)^2dt = \sum \frac{1}{2} m_i \vec
v_i^2 + \lim_{T \rightarrow \infty}~\frac{1}{T}\int^T_0~\frac{\sum \vec x_i \vec\nabla_iV}{2}~dt
\label{twentythree}
\eeq and hence we have an \underline{inequality} instead of an \underline{equality}.

In the quantum case, the average over \underline{time} of the kinetic energy is replaced by an average over \underline{space} of the
kinetic energy operator.  For the radial case the virial theorem can be proved by hand, but, following Thirring\cite{ff}, the best
general proof uses scaling.  Consider the Hamiltonian
\beq H = -\sum \frac{\Delta_i}{2m_i} + V(x_1, x_2,...,x_n)~,~~~~H \psi_k = E_k\psi_k
\label{twentyfour}
\eeq  
if we change the scale $x_i \rightarrow \lambda x_i$, the energy will be unchanged for
\beq H_\lambda = - \frac{1}{\lambda^2} \sum \frac{\Delta_i}{2m_i} + V(\lambda x_1,...,\lambda x_N)
\label{twentyfive}
\eeq 
so if we consider the Hamiltonian
\beq
\lambda^2H_\lambda = -\sum \frac{\Delta_i}{2m_i} + \lambda^2V(\lambda x_1,...,\lambda x_N)~,
\label{twentysix}
\eeq 
the energy levels $\epsilon_k$ are
\beq
\epsilon_k(\lambda) = \lambda^2E_k
\label{twentyseven}
\eeq 
so, for $\lambda = 1$, by the Feynman-Hellman theorem,
\beq
\frac{d\epsilon_k}{d\lambda} = 2 E_k = 2 \langle V(x_1,...,x_N)\rangle + \langle \Sigma \vec x_i \vec \nabla_iV(x_i,...,x_N)\rangle~,
\label{twentyeight}
\eeq so
$$ E_k = \langle V(x_1,...,x_N)\rangle + \frac{1}{2} \langle \Sigma \vec x_i \vec\nabla_i V \rangle
\nonumber
$$ 
and since $E_k = \langle T \rangle + \langle V \rangle$, $T$ being the kinetic energy operator,
\beq
\langle T \rangle = \frac{1}{2} \langle \Sigma \vec x_i \vec\nabla_i \rangle~.
\label{twentynine}
\eeq 
{\it Remark:  the choice of the origin is \underline{irrelevant} because $\langle \vec a \cdot \vec\nabla V \rangle = 0$ if $a$ is
a fixed vector. This can be verified in an elementary way, using the Schr\"odinger equation.}


\section{Virial theorem in the three-dimensional quantum case}

First, we would like to give also the ``pedestrian" proof of the virial theorem in the 3 dimensions radial case, to show the importance
of the prescription $u = 0$.

Take the Schr\"odinger equation
\beq -u" + Vu = Eu~~,
\label{thirty}
\eeq 
multiply by $u$ and integrate.  This gives
\beq -uu^\prime {\Bigg \vert}^\infty_0 + \int u^{\prime 2} dr + \int Vu^2dr = E\int u^2dr
\label{thirtyone}
\eeq 
multiply also by $2ru^\prime$ and integrate.  This gives
\bea -ru^{\prime 2} {\Bigg\vert}^\infty_0 &+& \int^\infty_0 u^{\prime 2} dr + rVu^2 {\Bigg\vert}^\infty_0 - \int^\infty _0 \left( Vu^2
+ r \frac{dV}{dr}u^2 \right) dr \nonumber \\ &=& Eru^2 {\Bigg\vert}^\infty_0  - E \int u^2 dr
\eea 
Assume that $u$ and $u^\prime$ vanish at $r \rightarrow \infty$.  Adding up the 2 equations, we get
\beq
\lim_{r \rightarrow 0} \left(uu^\prime + ru^{\prime 2} - rVu^2\right) + 2 \int u^{\prime 2} dr = \int^\infty_0 r \frac{dV}{dr}~u^2 dr
\label{thirtythree}
\eeq 
If $u(r) \sim r$ for $r \rightarrow 0$ and if $u^\prime (0)$ is finite, we get the virial theorem, if $u(0) = 1$ and if $u^\prime
(0)$ is not zero, the virial theorem is violated. If $ \int^r_{R_0} V(r^\prime )dr^\prime \rightarrow \pm \infty$ for $ r\rightarrow
0$, it is impossible to have $u^\prime (0) = 0$ because $\vert u^\prime (r) \vert \rightarrow \infty$ for $r \rightarrow 0$, and this
dominates $\lim rV$.  

If $V$ is integrable near the origin, one can save the virial theorem by taking $u^\prime (0) = 0$.  This, however, is equivalent to
looking at a symmetric one dimensional potential $V(x) = V(-x)$ and looking at the \underline{even} levels.  Notice that any other
boundary condition at the origin violates the virial theorem.

Now the point is that if the virial theorem holds at the classical level, it should hold too at the quantum level, according to the
sacro-sanct \underline{``correspondence principle"} of Bohr.

In the case of Coulomb interactions, we see that
$$
\frac{1}{2} r \frac{d}{dr}V(r) = -\frac{1}{2}~\frac{1}{r}
$$ 
so
$$
\langle T \rangle = - \frac{1}{2} \langle V \rangle = -E~.
$$ 
Taking a solution irregular at the origin for the case of Coulomb in \underline{3 dimensions} gives
$$
\langle V \rangle \sim - \int r^2 dr \frac{1}{r} \times \frac{1}{r^2} = -\infty
$$ 
and the virial theorem is completely violated.  The same is true for potentials more singular than Coulomb.

For the radial case, we have seen that we can save formally the virial theorem if $V$ is integrable at the origin and $u(0) = 1$
\underline{only if} $u^\prime (0) = 0$.  This corresponds to an \underline{unphysical} self-adjoint extension. Indeed while $\int
u^{\prime 2} dr$ is finite, $\int \vert \nabla \psi \vert^2d^2r$ diverges.

\section{Virial theorem in the two-dimensional quantum case}

In the radial case, the possible behaviour of the wave function at the origin are
\beq
\psi \sim c ~,  \psi \sim \ln r
\label{thirtyfour}
\eeq 
If $\psi \rightarrow c$, the kinetic energy is given by
\beq
\langle T \rangle = \int \Bigg\vert \vec\nabla \psi \Bigg\vert^2 dr^2 = -\int \psi \Delta\psi d^2r =
\frac{1}{2} \int \left( r \frac{dV}{dr}\right) \psi^2d^2r~,
\label{thirtyfive}\eeq 
and the virial theorem is satisfied.  More specifically, using the reduced wave function
\beq u = \sqrt{2\pi r}~~,
\label{thirtysix}
\eeq 
which satisfies
\beq -u'' - \frac{u}{4r^2} + Vu = Eu
\label{thirtyseven}
\eeq 
and using the same stragegy as in the three-dimensional case, we get
\beq
\langle T \rangle = \frac{1}{2} \lim_{r \rightarrow 0} \frac{|u(r)^2|}{r} + \int \left( u^{\prime 2} - \frac{u^2}{4r^2} \right) r dr  =
\frac{1}{2} \int r~ \frac{dV}{dr} u^2dr
\label{thirtyeight}
\eeq 
If on the other hand, $\psi \sim \ln r$ near the origin, 
$\nabla\psi \sim \frac{1}{r}$ 
and hence
\beq
\int \Bigg\vert \nabla \psi \Bigg\vert^2 rdr \sim \int \frac{dr}{r} = \infty
\label{thirtynine}
\eeq 
Therefore, \underline{in all cases}, for any potential, there is a violation of the virial theorem by the irregular solution. 
This, we believe, settles controversies on the choice of boundary conditions at the origin in a paper on the number of bound states in
one and two space dimensions\cite{ggg}.

\section{Acknowledgements}

I am grateful to Yves Cantelaube for a written exchange of ideas about the problem and to R. Balian for suggesting that I submit my
remarks for publication.


\begin{thebibliography}{99}
\bibitem{aaa} P.A.M. Dirac, "The principles of quantum mechanics", Oxford, Clarendon Press, 4th edition (1957) p. 156.
\bibitem{bb} L. Van Hove, Lectures at Les Houches Summer School, 1951, unpublished.
\bibitem{cc} C. Cohen-Tannoudji, B. Din and F. Lal\"oe, Quantum Mechanics, Hermann and John Wiley and sons, Paris-New York (1977), p.
782 and, in appendix IIp. 1476.
\bibitem{dd} J.L. Basdevant, M\'ecanique Quantique, Ecole Polytechnique, Editions Ellipses (1986) p. 185, J.L. Basdevant and J.
Dalibard, Springer Verlag, Berlin Heidelberg (2002), p. 217.
\bibitem{ee} See, for instance: M.L. Baekeman and K. Chadan, {\it Nucl. Phys.} {\bf A255} (1975) 35; and \\
Ann. Inst. Henri Poincar\'e XXIVA, 1-16.
\bibitem{ff} W. Thirring, Quantum Mechanics of Atoms and Molecules, Springer Verlag, New-York (1981) pp. 188 and 261 (here only the
special case of Coulomb forces is presented). 
\bibitem{ggg} K. Chadan, N.N. Khuri, A. Martin and T.T. Wu, {\it Journ. Math. Phys.}{\bf
44} (2003) 406.
\end{thebibliography}
\end{document}